\documentclass[journal, final, 12pt]{IEEEtran}
\usepackage[acronym]{glossaries}
\usepackage{graphicx}
\usepackage{xcolor}
\usepackage[binary-units=true]{siunitx}
\graphicspath{ {./figures/} }

\newacronym{SNR}{SNR}{signal-to-noise ratio}
\newacronym{LDPC}{LDPC}{low-density parity-check}
\newacronym{BP}{BP}{belief propagation}
\newacronym{QAM}{QAM}{quadrature amplitude modulation}
\newacronym{QPSK}{QPSK}{quadrature phase-shift keying}
\newacronym{BICM}{BICM}{bit-interleaved coded modulation}
\newacronym{BCE}{BCE}{binary cross-entropy}
\newacronym{LLR}{LLR}{log-likelihood ratio}
\newacronym{BMI}{BMI}{bit-wise mutual information}
\newacronym{KL}{KL}{Kullback–Leibler}
\newacronym{BMD}{BMD}{bit-metric decoding}
\newacronym{BER}{BER}{bit error rate}
\newacronym{NN}{NN}{neural network}
\newacronym{GS}{GS}{geometric shaping}
\newacronym{SIP}{SIP}{superimposed pilot}
\newacronym{iid}{i.i.d.\@}{independent and identically distributed}
\newacronym{SGD}{SGD}{stochastic gradient descent}
\newacronym{wrt}{w.r.t.\@}{with respect to}
\newacronym{MAP}{MAP}{maximum a posteriori}
\newacronym{LMMSE}{LMMSE}{linear minimum mean square error}
\newacronym{AWGN}{AWGN}{additive white Gaussian noise}
\newacronym{RBF}{RBF}{Rayleigh block fading}
\newacronym{OFDM}{OFDM}{orthogonal frequency division multiplexing}
\newacronym{3GPP}{3GPP}{3rd Generation Partnership Project}
\newacronym{5GNR}{5G NR}{5G New Radio}
\newacronym{PRB}{PRB}{physical resource block}
\newacronym{IEDD}{IEDD}{iterative estimation, demapping, and decoding}
\newacronym{CSI}{CSI}{channel state information}
\newacronym{MSE}{MSE}{mean squared error}
\newacronym{BPSK}{BPSK}{binary phase-shift keying}
\newacronym{DC}{DC}{direct current}
\newacronym{PAPR}{PAPR}{peak-to-average power ratio}
\newacronym{DMRS}{DMRS}{demodulation reference signal}
\newacronym{TTI}{TTI}{transmission time interval}
\newacronym{CDF}{CDF}{cumulative distribution function}
\newacronym{RE}{RE}{resource element}
\newacronym{MIMO}{MIMO}{multiple-input multiple-output}
\newacronym{ML}{ML}{machine learning}
\newacronym{AI}{AI}{artificial intelligence}
\newacronym{PHY}{PHY}{physical}
\newacronym{MAC}{MAC}{medium access control}
\newacronym{ADC}{ADC}{analog-to-digital converter}
\newacronym{DAC}{DAC}{digital-to-analog converter}
\newacronym{RAN}{RAN}{radio access network}
\newacronym{PRACH}{PRACH}{physical random access channel}
\newacronym{ack}{ACK}{acknowledgement}
\newacronym{api}{API}{application programming interface}
\newacronym{bs}{BS}{base station}
\newacronym{bsr}{BSR}{buffer status report}
\newacronym{drx}{DRX}{discontinuous reception}
\newacronym{ic}{IC}{instantaneous coordination}
\newacronym{l2c}{L2C}{learning to communicate}
\newacronym{marl}{MARL}{multiagent reinforcement learning}
\newacronym{phr}{PHR}{power headroom report}
\newacronym{rl}{RL}{reinforcement learning}
\newacronym{sdu}{SDU}{service data unit}
\newacronym{sg}{SG}{scheduling grant}
\newacronym{sr}{SR}{scheduling request}
\newacronym{ta}{TA}{timing advance}
\newacronym{ts}{TS}{technical specification}
\newacronym{ue}{UE}{user equipment}
\newacronym{SISO}{SISO}{single-input single-output}

\begin{document}

\title{Toward a 6G AI-Native Air Interface}

\author{Jakob Hoydis,~\IEEEmembership{Senior Member, IEEE}, Fay\c{c}al Ait Aoudia,~\IEEEmembership{Member, IEEE},\\ Alvaro Valcarce,~\IEEEmembership{Senior Member, IEEE}, and Harish Viswanathan,~\IEEEmembership{Fellow, IEEE}
\thanks{The authors are with Nokia Bell Labs (\{jakob.hoydis, faycal.ait\_aoudia, alvaro.valcarce\_rial, harish.viswanathan\}@nokia-bell-labs.com).}%
}%

\maketitle

\begin{abstract}
Each generation of cellular communication systems is marked by a defining disruptive technology of its time, such as \gls{OFDM} for 4G or Massive \gls{MIMO} for 5G. Since \gls{AI} is the defining technology of our time, it is natural to ask what role it could play for 6G. While it is clear that 6G must cater to the needs of large distributed learning systems, it is less certain if \gls{AI} will play a defining role in the design of 6G itself. The goal of this article is to paint a vision of a new air interface which is partially designed by \gls{AI} to enable optimized communication schemes for any hardware, radio environment, and application.
\end{abstract}

\section{Introduction}
While 5G is rolled out globally and the standardization discussions for its future evolution take place, researchers in academia and industry start to think about visions, use cases, and disruptive key technologies for a possible 6G system. Publicly funded 6G research projects in Europe \cite{6gwaves2020}, the United States, and China are under way, and also the ITU has begun their work on requirements for fixed networks in the 2030s \cite{network2030}. A common theme in many 6G vision papers is that of creating digital twin worlds for seamlessly connecting and controlling physical and biological entities to enable new  mixed-reality super-physical experiences \cite{viswanathan2020}. 

Apart from new spectrum technologies and the support of simultaneous communications and sensing as well as extreme connectivity requirements (among others), it is expected that \gls{ML} and \gls{AI} will play a defining role in the development of 6G networks end-to-end across the design, deployment, and operational phases \cite{letaief2019roadmap}. 
As the network evolves to programmable and flexible cloud native implementation, \gls{ML}/\gls{AI}-based network automation will be crucial to simplify network management and optimization. Networks will become ``cognitive'' in the sense that various aspects such as virtualized network function placement, slicing, quality of service, mobility management, radio resource management, and spectrum sharing  will all rely on ML/AI to varying degrees. In fact, we expect that ML/AI will significantly impact even the 6G air interface, which is the focus of this paper.

While \gls{ML} starts to be widely used in the industry to enhance the implementation of various components within the 5G \gls{RAN} and core, it is fair to say that there is not a single component of 5G which has been \emph	{designed by} \gls{ML}. The purpose of this article is therefore to raise and discuss the question: What if 6G was designed in a way that \gls{ML}/\gls{AI} could modify parts of the \gls{PHY} and \gls{MAC} layers?

\begin{figure*}
\begin{center}
\includegraphics[width=0.9\textwidth]{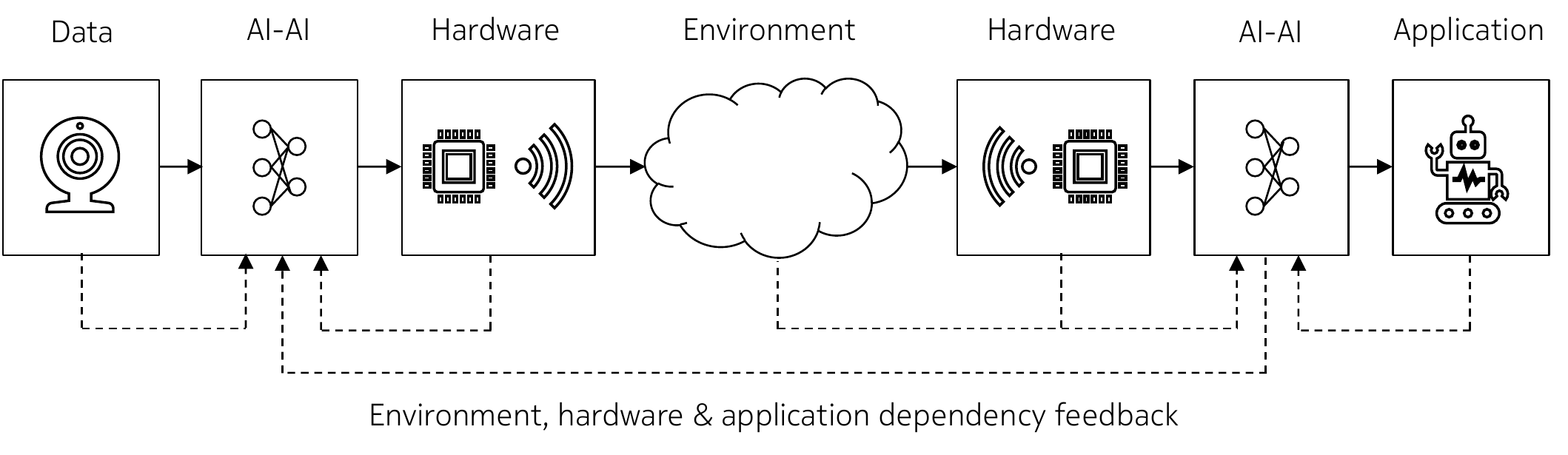}
\caption{The AI-Native Air Interface (AI-AI) adapts to different radio environments, hardware, data, and applications. Compared to previous air interfaces, it is not only designed to reliably transmit bits, but also to serve an application with the data it needs in an optimal way.}
\label{fig:aiai}
\end{center}
\end{figure*}

\section{AI-Native Air Interface}
Let us start by presenting our vision and motivation of an \emph{AI-Native Air Interface (AI-AI)} which describes a disruptive change to the traditional way communication systems are designed, standardized, and productized. We will first provide a summary of possible benefits of such an approach, then detail three important but also necessary development steps towards realizing our vision, and finally  present a case study which exemplifies the respective potential performance gains and advantages. 

As illustrated in Fig.~\ref{fig:aiai}, the goal of the AI-AI is to serve an application with the data it needs in the most efficient way by taking into account the constraints of the available hardware and the radio environment. The AI-AI hence no longer decouples source and channel coding as well as communication of data from the intended use by an application, and embraces hardware constraints and undesired effects of the communications channel rather than fighting them. While the last decades were used to implement the scientific breakthroughs by Shannon and Wiener (as well as many others), we are now entering a new era for communications where classical approaches must be revisited and new theories developed to achieve the technological breakthroughs needed for a possible 6G system.
We believe that our vision of an AI-AI could become a reality within the next decade by optimizing the air interface from end-to-end thanks to advances in the field of \gls{AI} for communications. 

\subsection{Possible benefits}
First, in contrast to a single classical waveform choice such as \gls{OFDM} in 5G, the AI-AI could enable learning of bespoke waveforms for different frequencies which do not only make more efficient use of the spectrum but are also optimally adapted to practical limitations of the transceiver hardware and channel, such as non-linear power amplifiers, hybrid analog-digital processing, low quantization resolution, very short channel coherence time and bandwidth, phase and impulsive noise. Also new modulation schemes, pilot sequences, and codes can be learned or optimized with \gls{ML} to squeeze even more performance out of the spectrum. Several research groups have recently demonstrated practical gains of such approaches, see, e.g., \cite{cammerer2020, downey2020}. A very interesting application arises for the transmission of short messages, where the classical frame structure of preamble, pilots, coded bits, and cyclic-redundancy check could be replaced with fully learned radio burst conveying a few bits of information. Furthermore, learning of new waveforms for simultaneous communication + ``X'' (e.g., sensing or power transfer) has high potential.

Second, fully learned transceivers have the benefit that they do not need to undergo the very costly and time-consuming traditional process of algorithm design and hardware implementation anymore. They can be trained directly for the targeted hardware platform (or even on it, depending on the capabilities). With an increasingly rich diversity of expected 6G use-cases and the emergence of small-scale \emph{sub-networks} \cite{ziegler20206g}, this versatility becomes essential to ensure that 6G can cater for each individual use-case and deployment scenario in the best way possible. Moreover, given the rapid speed at which \gls{ML} hardware accelerators are developing, it is likely that learned transceiver implementations will rapidly outperform their traditional counterparts in power efficiency, latency, and cost. Advances in neuromorphic computing \cite{rajendran2019low} could further amplify this trend.

Third, the more we follow the AI-AI principle of learning based design and specification, the less needs to be standardized. The current 5G specification boasts a very rich sets of options and parameters for different frequency bands and scenarios which pose a difficult challenge from an implementation point of view. It is undesirable to scale this approach to even more complex and diversified settings in 6G. If, on the other hand, only a sufficiently flexible framework for air interface learning was standardized, the system could auto-adjust to any kind of scenario. With a bit of wishful thinking, one could hope that 6G could be the last communication system to be standardized.

\begin{figure*}
\begin{center}
\includegraphics[width=0.9\textwidth]{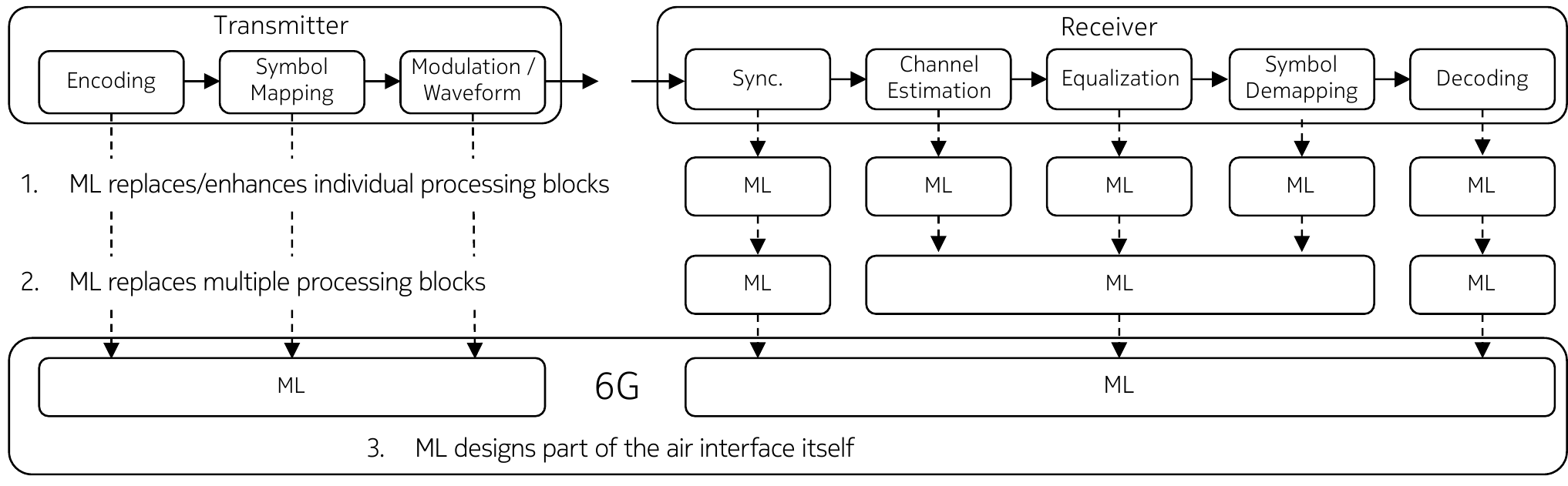}
\caption{Three phases towards the AI-AI: After gradually replacing most of the processing blocks, \gls{ML} will design parts of the \gls{PHY} layer.}
\label{fig:3steps}
\end{center}
\end{figure*} 

Fourth, the AI-AI allows integration of the data and the application consuming it into a single end-to-end learning process. Using the terminology from Shannon and Weaver's seminal book \cite{shannonweaver}, the AI-AI no longer only solves the problem of reliably transmitting bits (Level A), but simultaneously addresses the problems of semantics (Level B) and effectiveness (Level C) of  communication (see, e.g., \cite{popovski2020semantic}). 
While the latter aspects  may not be applicable to the generic Internet communication scenario, they become relevant for communication systems which are tailored to specific purposes and under the control of a single entity, such as industrial communication systems for sensing, surveillance, and robot control.

Lastly, the idea of end-to-end learning naturally extends to the \gls{MAC} layer where it would be desirable to emerge optimized signaling schemes and channel access policies which fluently transition from contention- to schedule-based depending on the use-case and environment. Protocol learning could also address the problem of optimally multiplexing resources for communication and sensing (or other applications that radio waves can be used for). Ultimately, \gls{PHY} and \gls{MAC} layers could be jointly learned together.

\subsection{Three steps towards the AI-AI}
We currently see three important phases in the development and transition to the AI-AI, each of which requires sustained multi-disciplinary research. These are schematically shown in Fig.~\ref{fig:3steps}. The first two phases do not require any new signaling or procedures as they only impact the implementation of transceiver algorithms. They can therefore be carried out on future 5G systems to gather practical experience while the research on 6G is progressing.  

\subsubsection{\gls{ML} replaces single processing blocks}
In the first phase, which is already happening in the industry today, \gls{ML} will be used to enhance or replace some of the processing blocks, mostly in the receiver. Examples are physical random access channel detection, channel estimation, or symbol demapping.  Although seemingly simple, this step constitutes a paradigm change in the way the industry designs and deploys radio transceivers. Even if the \gls{ML} models are likely to be rather small, several important problems such as data acquisition, model updates, and online training need to be solved and hardware accelerators must be integrated into the \gls{PHY} processing flow. The receiver processing will contain a mix of \gls{ML} and traditional blocks.

\subsubsection{\gls{ML} replaces multiple processing blocks}
In the second phase, more functionality is given to \gls{ML} models which take on the joint role of multiple processing blocks. This could be, e.g., joint channel estimation, equalization, and demapping. In this phase of the transition, the \gls{ML} models will grow larger, hardware acceleration becomes increasingly important, and vendors need to commit to an `ML-only/ML-first' approach because it is not viable to implement ML and non-ML backup solutions in parallel in the same processing platform due to increased power consumption and cost. This means that ML is also trusted more although the inner workings of large models are less interpretable, but the potential gains are also higher. An example will be provided in Section~\ref{sec:case-study}. In this phase, we will also realize what possibilities such learned transceiver components open up, e.g., need for less pilots, no cyclic prefix, less stringent synchronization. In other words, we will learn what are the things \gls{ML} allows us to do which we could not do before (with reasonable effort). 

\subsubsection{\gls{ML} designs parts of the air interface}
In the third phase, we will give even more freedom to \gls{ML}/\gls{AI} and let it design parts of the physical and \gls{MAC} layers itself. This represents another paradigm change in the way communication systems are designed because not all aspects of the \gls{PHY} and \gls{MAC} layers might be fixed in advance. This approach requires new forms of signaling and procedures to enable distributed end-to-end training. Rather than specifying, e.g., modulation schemes and waveforms, one would need to specify procedures that can be used to optimize these aspects of the air interface at deployment time.  
This is clearly something nobody has ever done before and that requires a massive change in the way communication systems are standardized. It is of course also possible that ML-designed solutions to specific problems will be specified, which is, e.g., already the case in 5G for the channel code design.

\begin{figure}
\begin{center}
\includegraphics[width=0.95\columnwidth]{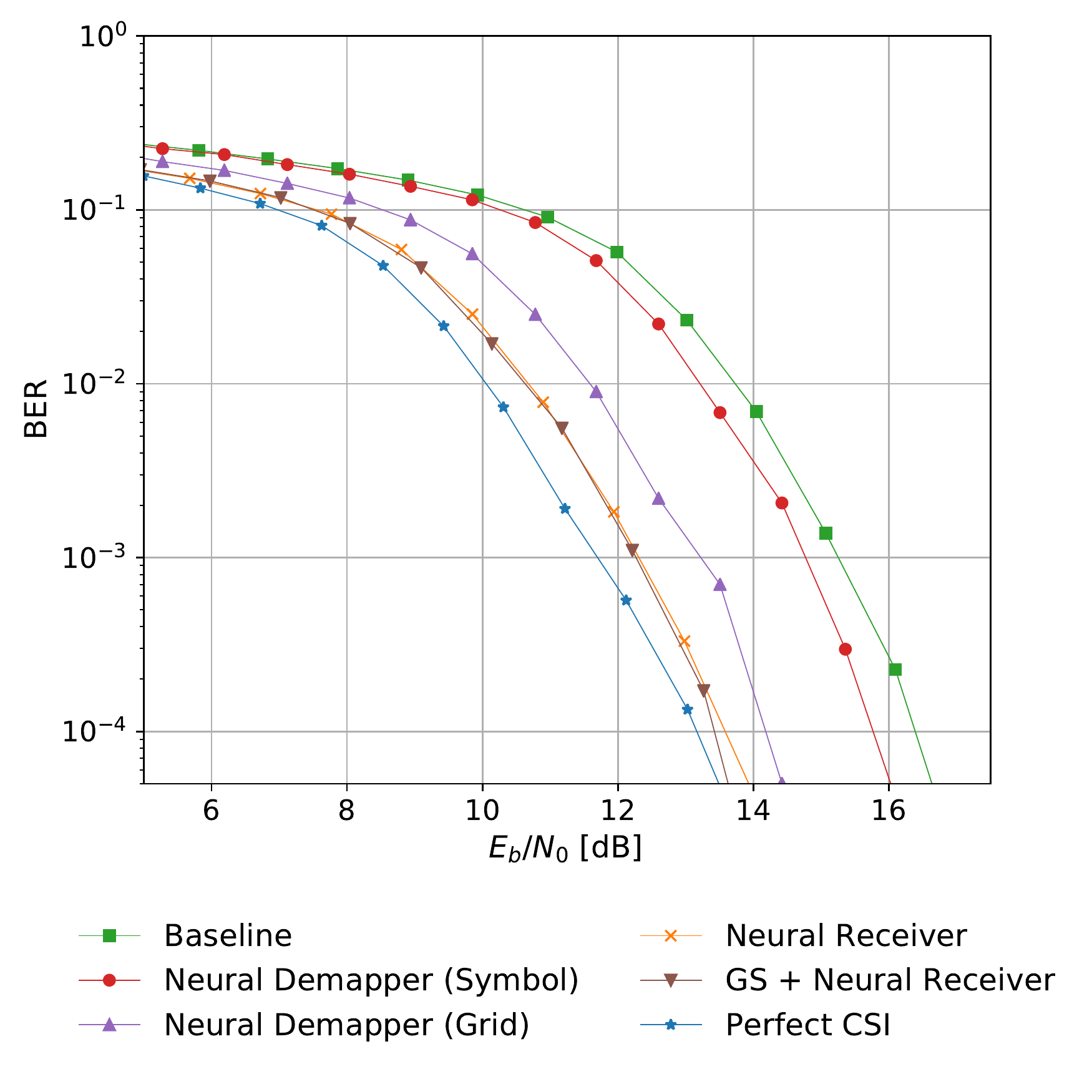}
\caption{\gls{BER} performance of all compared schemes in the case study.}
\label{fig:ber}
\end{center}
\end{figure}
\vspace{-10pt}

\subsection{Case study: From neural receivers to pilotless transmissions}
\label{sec:case-study}
Next, we will present a case study that illustrates the progression through the three phases towards the AI-AI outlined above and demonstrates the respective performance gains. We consider a doubly-selective \gls{SISO} channel at a carrier frequency of \SI{3.5}{\giga\hertz} with the TDL-A power delay profile  and a delay spread of \SI{100}{\nano\second}. The receiver is assumed to move at a speed of \SI{50}{\kilo\meter\per\hour} and the channel evolves in time according to Jakes' model. We consider cyclic prefix-based \gls{OFDM} with \num{72} subcarriers spaced \SI{30}{\kilo\hertz} apart and assume  \glspl{TTI} of \num{14} consecutive \gls{OFDM} symbols which contain codewords of length \SI{1024}{\bit} at a coderate of \num[fraction-function = \sfrac]{2/3}, generated by a 5G-compliant code.

Our non-ML baseline assumes \num{64}-\gls{QAM}, pilots transmitted on every other sub-carrier on the third and twelfth \gls{OFDM} symbols, least-squares channel estimation, equalization based on the nearest pilot, exact demapping to \glspl{LLR} assuming a Gaussian post-equalized channel, as well as a standard \gls{BP} decoder. 

The \gls{BER} performance of the baseline and all other schemes that will be subsequently introduced is shown in Fig.~\ref{fig:ber}. One can see that there is approximately a \SI{3}{\decibel} gap between the  baseline and a receiver assuming perfect \gls{CSI}. We now describe some ways to close this gap using \gls{ML}-enhanced receiver processing before delving into the benefits of optimizing parts of the transmitter, too.   

Due to channel aging and imperfect channel estimation, the quality of the post-equalized symbols which are fed into the demapper changes over the grid of \glspl{RE} within a \gls{TTI}. A first possibility to cope with this problem is to learn a bespoke neural demapper for each \gls{RE} (Phase 1). The \gls{BER} performance of such a scheme is shown by the red line with dot markers in Fig.~\ref{fig:ber}. As expected, it provides some \SI{0.5}{\decibel} gain over the baseline by computing better \glspl{LLR}, but cannot compensate for channel aging which results in a rotation and scaling of the equalized constellation. 

\begin{figure}
\begin{center}
\includegraphics[width=0.9\columnwidth]{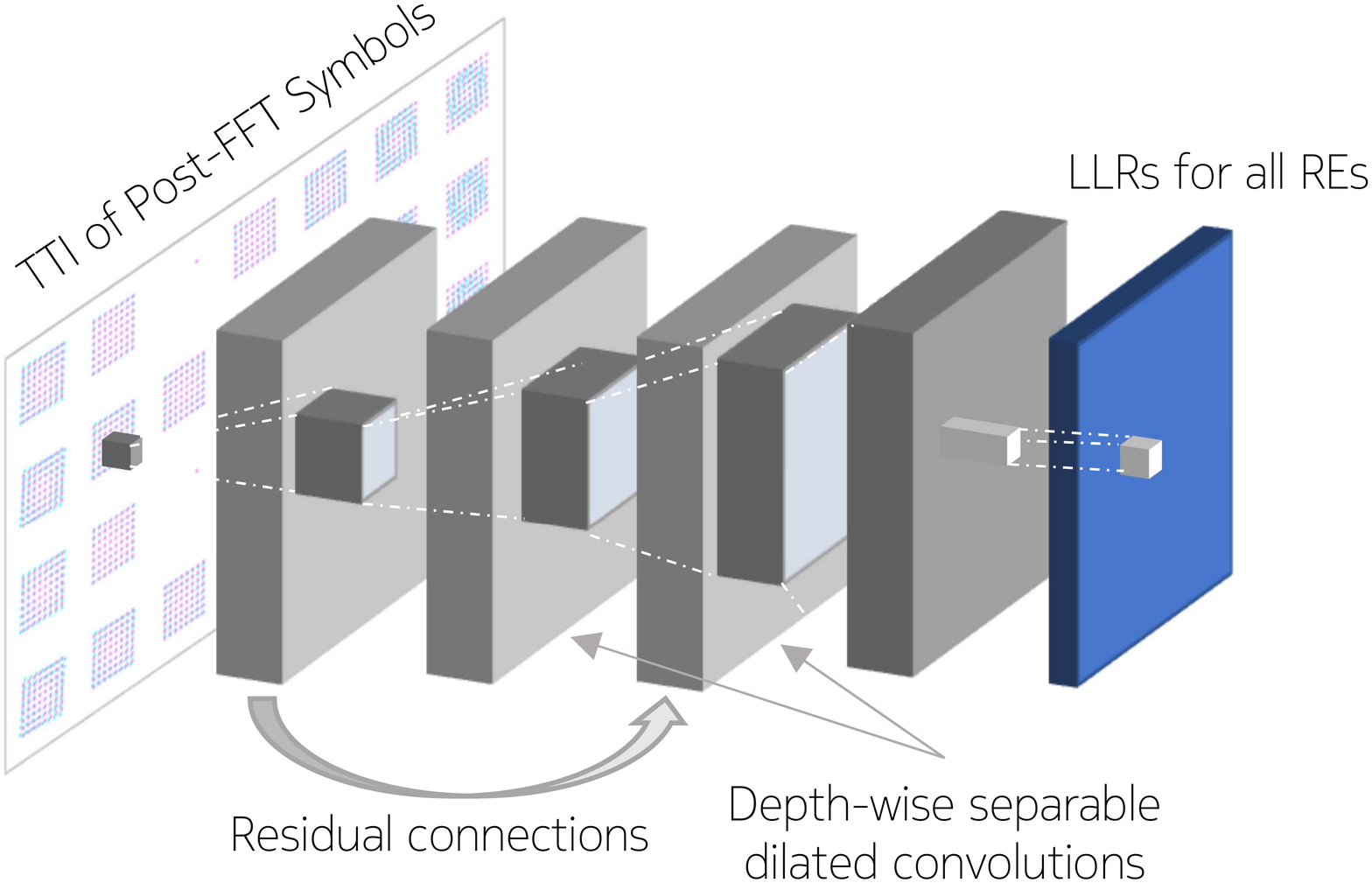}
\caption{The neural receiver produces \glspl{LLR} for an entire \gls{TTI} of post-FFT symbols. Key architectural components are depth-wise separable dilated convolutions and residual connections. The same architecture can also be used as a neural demapper, operating on a \gls{TTI} of equalized symbols.}
\label{fig:resnet}
\end{center}
\end{figure}

In order to address these shortcomings, one can use a larger neural demapper which does not operate symbol-by-symbol but rather produces \glspl{LLR} for the full \gls{TTI}. It was shown in \cite{honkala2020, aitaoudia2020}, that a fully convolutional ResNet architecture with dilated separable convolutions achieves remarkable performance for this task (see Fig.~\ref{fig:resnet}). By having access to the full \gls{TTI} of post-equalized symbols, the learned demapper can compensate for some of the errors made by the channel estimator and equalizer to provide a \SI{2}{\decibel} improvement over the baseline (see purple line with triangular markers in Fig.~\ref{fig:ber}). 

Interestingly, it turns out that one can assign the joint task of channel estimation, equalization, and demapping to a neural network with a similar architecture (Phase 2). It is fed with a \gls{TTI} of post-FFT received signals from which it directly computes \glspl{LLR} for all bits. In addition to the gains of the  learned demapper, this neural receiver is now able to carry-out data-aided channel estimation and detection, resulting in an additional \SI{0.5}{\decibel} gain. By increasing the model complexity and the size of the input (more sub-carriers and \gls{OFDM} symbols), the performance can be brought arbitrarily close to the perfect \gls{CSI} performance \cite{honkala2020}.

\begin{figure}
\begin{center}
\includegraphics[width=0.5\columnwidth]{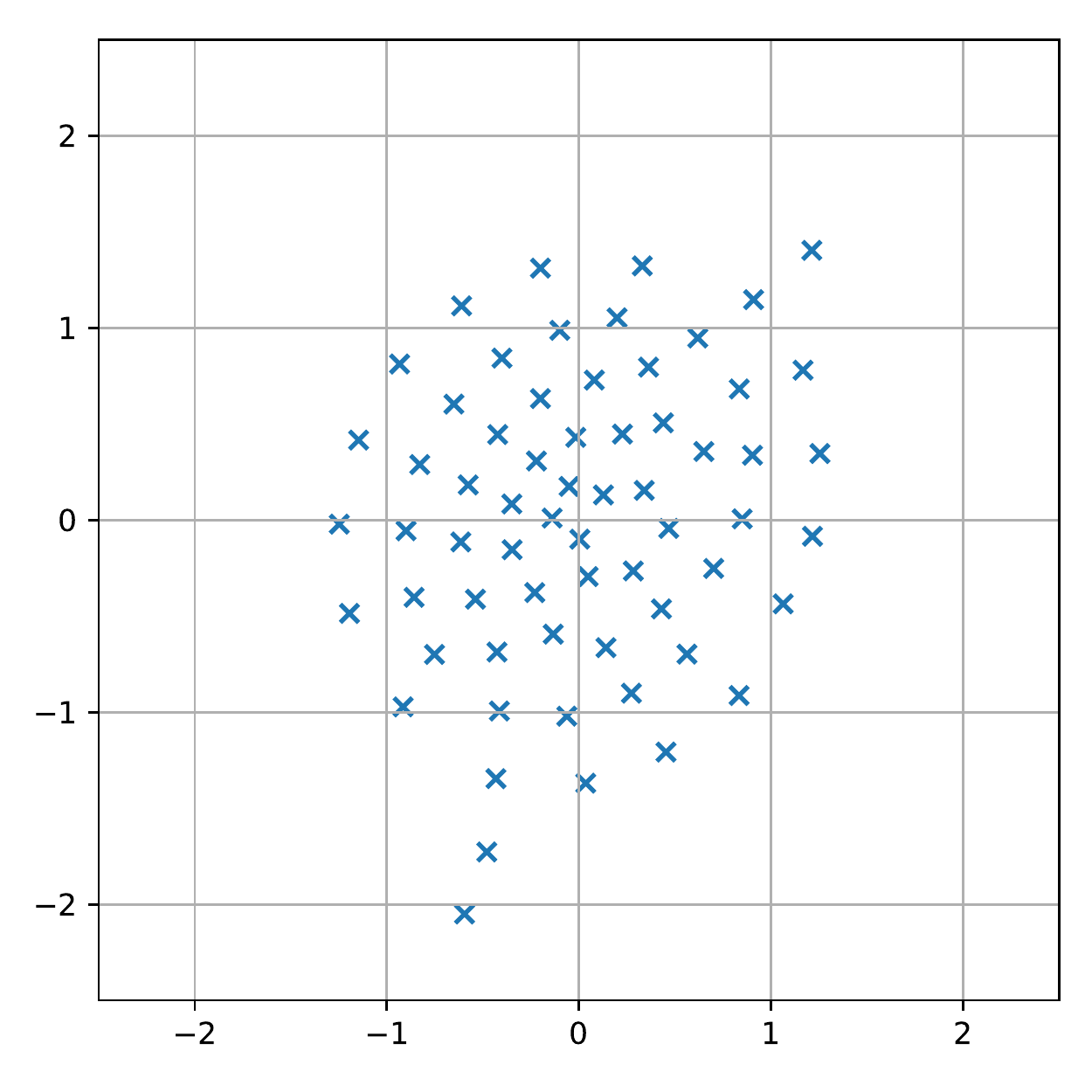}
\caption{Learned constellation allowing pilotless transmissions together with a neural receiver. It has zero mean, unit power, and a single axis of symmetry. The optimal bit-labeling is also learned but not shown for readability.}
\label{fig:const}
\end{center}
\end{figure}

Lastly, we would like to investigate the benefits of a learned constellation (i.e., geometric shaping (GS)) at the transmitter side (Phase 3), which is jointly optimized together with the neural receiver. Fig.~\ref{fig:const} shows this constellation which is used on every \gls{RE} instead of the mix of pilots and \num{64}-\gls{QAM} symbols sent by the baseline. As can be seen from Fig.~\ref{fig:ber}, this system achieves the same \gls{BER} as the neural receiver with \num{64}-\gls{QAM}, but has the additional benefit that no pilots are transmitted. End-to-end learning could hence remove the need and control overhead for demodulation reference signals.

This case study has only scratched the surface of what will be possible in the future. Interesting directions for future research include end-to-end learning for new waveforms, constrained hardware, very short messages, as well as joint source-channel coding for a specific application (which is also learned). Meta, transfer, and federated learning are key enablers to make such schemes practical.

\section{The next frontier: Protocol learning for the \gls{MAC}}
The protocols above the \gls{PHY} take bit-by-bit transmission for granted to develop complex signaling schemes and orchestrate elaborate procedures across the network's nodes.
As a result, radios coordinate harmoniously and provide more powerful services than what point-to-point links offer.

The telecommunications industry defines these procedures through protocol standards which are agreed upon in large meetings, where competing technical and economical interests are debated year after year.
These efforts have a high cost and sometimes result in ambiguous \glspl{ts}.
After a \gls{ts} is released, the implementation and test phase begins, which is even more costly. For this reason, it is interesting to question if this burdensome undertaking could be somehow automated. And if so, would the result perform better than human-designed protocols?

Wireless protocols are sequences of messages exchanged between radio nodes to transmit \glspl{sdu}. As such, protocols can also be understood as a language between collaborative machines. Learning a language is something not only humans but also machines can do \cite{Brown2020}.
In fact, the field of \gls{l2c} is growing fast thanks to recent developments in deep \gls{marl}, see, e.g., \cite{Foerster2016}. While most research efforts in this field are targeted towards natural languages, we believe these techniques can also be used for training wireless devices to learn communication protocols. The 6G protocols of the AI-AI could be built this way and there are two ways to achieve this:

\subsection{Learning a given protocol}
Intelligent software agents could be trained to behave according to an a priori known protocol.
Instead of coding the standard, agents could be trained on it via machine learning.
Such training would ideally be done only once during factory production.
This would yield a protocol implementation and, although it does not replace protocol standardization, could replace protocol interpretation, implementation, and testing efforts. The cost-savings and time-to-market reduction potential of 6G nodes built this way might be significant.

Already today, \glspl{ue} could be trained to learn the \gls{5GNR} \gls{MAC} protocol.
This would include learning to interpret the different control messages received from the \gls{bs} (e.g., \gls{drx}, \gls{ta}, etc.), as well as learning what to send in the uplink (e.g., \gls{bsr}, \gls{phr}, etc.). Learning can be formulated as a \gls{marl} problem, wherein the \glspl{ue}' \glspl{MAC} are deep \gls{rl} agents with two action spaces:
\begin{itemize}
	\item \emph{Uplink signaling action space}: All possible uplink control messages a \gls{ue} may ever send.
	\item \emph{\gls{PHY} action space}: All channel access commands the \gls{MAC} may invoke through the \gls{PHY} \gls{api}.
\end{itemize}

\begin{figure}
\begin{center}
\includegraphics[width=0.9\columnwidth]{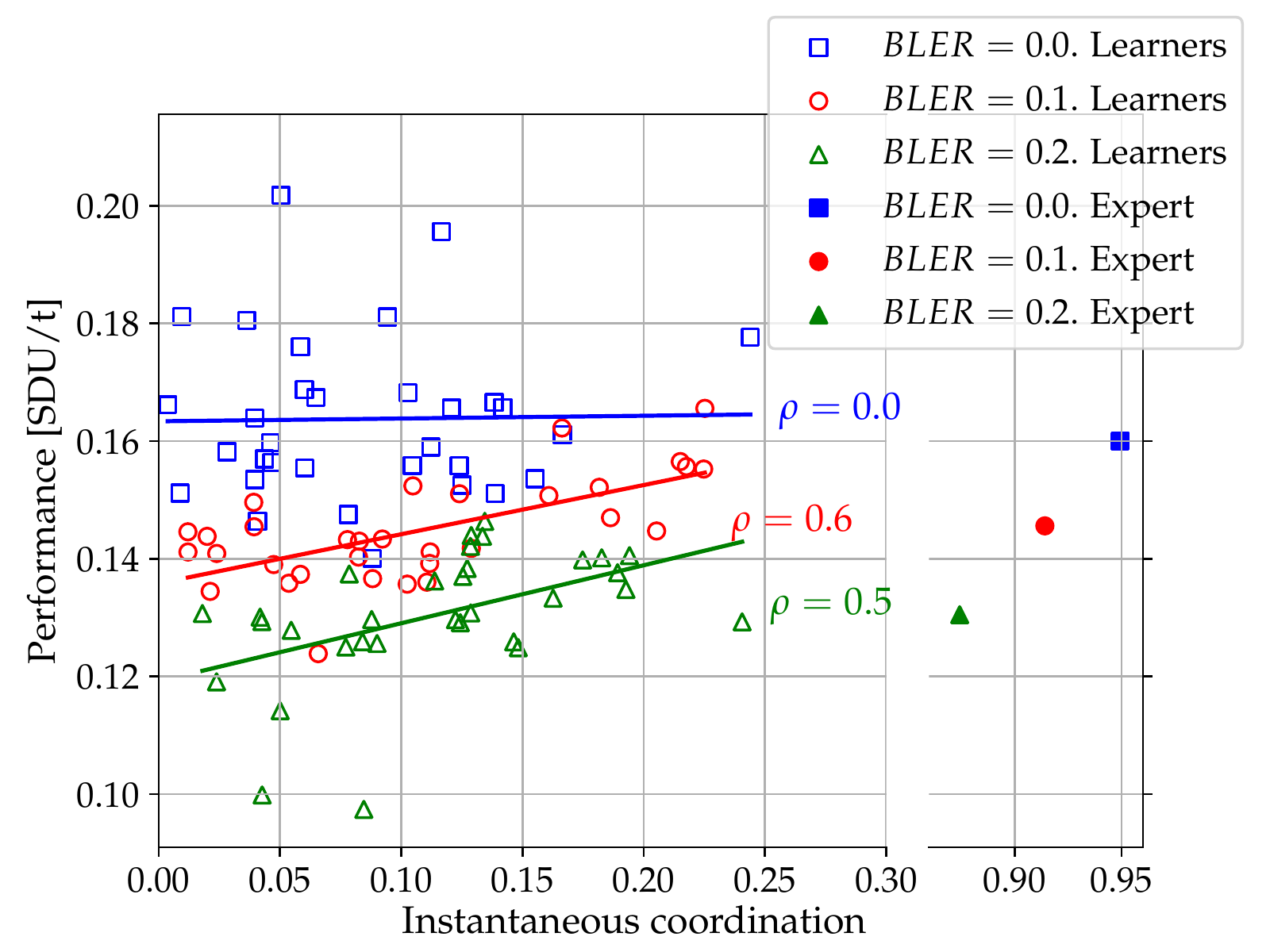}
\caption{Relationship between performance and signaling-based coordination. Each point depicts the mean \gls{sdu} rate and the \gls{ic} of different learned \gls{ue} \glspl{MAC}. Two \glspl{ue} were trained to transmit two randomly arriving \glspl{sdu} to the \gls{bs} without collision or channel loss, via a protocol known to the \gls{bs} and unknown to the \glspl{ue} (i.e., \glspl{ue} had no mapping between the protocol's messages and \gls{PHY} actions).}
\label{fig:ic}
\end{center}
\end{figure}

Protocol implementations trained this way may outperform expert systems thanks to the customized signaling and channel access policy \cite{Valcarce2020}. The signaling is the vocabulary of messages the nodes have at their disposal as well as the rules about how to use them. The channel access policy decides how to make use of the \gls{PHY} \gls{api}, based on information it has thanks to the signaling. It is therefore fair to ask whether the gains are due to the learned signaling, the channel access policy, or both. 
The impact of signaling on the \glspl{ue}' actions can be quantified by metrics such as the \gls{ic}, which is the mutual information between the downlink signaling messages and the next channel access actions.
This is illustrated in Fig.~\ref{fig:ic}, which shows that in unreliable channels, higher levels of coordination lead to performance gains (note the positive Pearson correlation coefficients $\rho$). The semantics of the \gls{MAC} messages used in Fig.~\ref{fig:ic} are fixed and known to the \gls{bs}.
Nevertheless, \gls{ML} training can yield \glspl{ue} with widely different interpretations of these messages and consequently, also very different policies and performance (note the high variance of the results in Fig.~\ref{fig:ic}).
This variance is a consequence of the vast size of the solution space, which strengthens the case for optimizing the protocols our radios use.

\vspace{-10pt}

\subsection{Emerging a new protocol}
As there are many possible ways a \gls{ue} can communicate with a \gls{bs} that implements a given protocol, why should we constrain ourselves to human-designed protocols? The next frontier in protocol learning is to let \glspl{ue} and \glspl{bs} explore the entire space of possible protocols. This is challenging because for two radios to coordinate, they first need to find a state where they can interpret each other's messages.
Recent \gls{l2c} research suggests that this language discovery problem may be overcome by first training the nodes with supervised learning.
This essentially endows radios with an initial protocol that they can later evolve through self-play.

Communication protocols emerged this way may be hard to interpret, which is essential for fault detection or performance monitoring.
For this reason, some use-cases may favor protocols that are close to known ones.
This requires metrics that measure the \emph{distance} between two protocols. Training techniques minimizing this distance may improve intelligibility.

The ability to learn new communication protocols opens the door to radio systems that are highly tailored to their deployment environment, thus boosting 6G capacities for niche and vertical markets. The AI-AI will not only reduce today's signaling overheads, but also the standardization and development efforts for the highly complex radio technologies of the next decades. We foresee a 6G future where parts of the radio stack development cycle could be replaced by the click of a button.

\section{Conclusion}
While the next decade will reveal if our vision of an AI-Native Air Interface provides sufficiently compelling benefits to make it into 6G, we are certain that AI/ML will profoundly change the way communication systems will be designed and deployed in the future. We hope that some of the readers will join us on this exciting journey.

\subsection*{Acknowledgments}
We would like to thank P. Mogensen, S. ten Brink, S. Cammerer, S. Dörner, M. Goutay, P. Srinath, M. Honkala, D. Korpi, J. Huttunen (and many others) for numerous discussions that helped to shape the vision outlined in this article.

\bibliographystyle{IEEEtran}

\begin{IEEEbiographynophoto}
{Jakob Hoydis} [SM'19] is head of a research department at Nokia Bell Labs, France, focusing on radio systems and artificial intelligence. He received his Ph.D.\ degree from Supélec, Gif-sur-Yvette, France, in 2012, He is a co-author of the textbook “Massive MIMO Networks: Spectral, Energy, and Hardware Efficiency” (2017). He is currently chair of the \textit{IEEE COMSOC Emerging Technology Initiative on Machine Learning for Communications} as well as Editor of the \textit{IEEE Transactions on Wireless Communications}.
\end{IEEEbiographynophoto}
\vspace{-15pt}

\begin{IEEEbiographynophoto}
{Fayçal Ait Aoudia} [M'20] is a research engineer at Nokia Bell Labs, France, where he is working on machine learning for wireless communications. He received the M.Sc. degree in computer science engineering from INSA Lyon, France, in 2014 and the Ph.D. degree from the University of Rennes 1, France, in 2017. His Ph.D. work has focused on energy management and communication protocol design for autonomous wireless sensor networks. 
\end{IEEEbiographynophoto}
\vspace{-15pt}

\begin{IEEEbiographynophoto}
{Alvaro Valcarce} [SM'20] is a research engineer at Nokia Bell Labs, France, where he focuses on the application of reinforcement learning to L2 and L3 problems for the development of beyond-5G technologies. He was previously a system engineer with Node-H GmbH, where he developed Self Organising Networks (SON) algorithms for LTE small-cells. He received his PhD in 2010 from the University of Bedforshire and his background is on cellular networks, optimization algorithms, computational electromagnetics, and satcom.
\end{IEEEbiographynophoto}
\vspace{-15pt}

\begin{IEEEbiographynophoto}
{Harish Viswanathan} [F’13] is head of the Radio Systems Research group at Nokia Bell Labs, USA, and a Bell Labs Fellow. He received his B.Tech. from Electrical Engineering, IIT Madras, Chennai, India, and the M.S. and Ph.D. from Electrical Engineering, Cornell University, Ithaca, NY, USA. Since joining Bell Labs in 1997, he has worked extensively on wireless research, ranging from physical layer to network architecture and protocols. From 2012 to 2015, he was a CTO Partner in Alcatel-Lucent, advising the Corporate CTO on technology strategy.
\end{IEEEbiographynophoto}
\end{document}